\documentclass[fleqn]{2020SCGE}
\usepackage{graphicx}
\usepackage{multirow}
\usepackage{hyperref}
\newcommand{\msunh}{\>h^{-1}\rm M_\odot}

\newcommand{\kms}{\>{\rm km}\,{\rm s}^{-1}}

\newcommand{\rmag}{\>^{0.1}{\rm M}_r-5\log h}

\begin{document}

\ensubject{subject}

%%%%%%%%%%%%%%%%%%%%%%%%%%%%%%%%%%%%%%%%%%%%%%%%%%%%%%%
%%% Authors do not modify the information below
%%% ????????????????
%%% ??????????, ????????????{}, ???????????????????
%Letter to the Editor??Article%??????
\ArticleType{Article}%??Article
\SpecialTopic{SPECIAL TOPIC: }%???????
\Year{2023}
\Month{January}
\Vol{66}
\No{1}
\DOI{??}
\ArtNo{000000}
\ReceiveDate{January 11, 2023}
\AcceptDate{April 6, 2023}
%\OnlineDate{January 1, 2016}
%%%%%%%%%%%%%%%%%%%%%%%%%%%%%%%%%%%%%%%%%%%%%%%%%%%%%%%

%%% title: ????
%%%   \title{title}{title for citation}
\title{Establishing HI mass v.s. stellar mass and halo mass scaling relations using an abundance matching method} 

%%% Corresponding author: ???????
%%%   \author[number]{Full name}{{email@xxx.com}}
%%% General author: ???????
%%%   \author[number]{Full name}{}

\author[1]{Yi Lu}{{luyi@shao.ac.cn}}%
\author[2,3]{Xiaohu Yang}{xyang@sjtu.edu.cn}
\author[3]{Chengze Liu}{}
\author[1]{Haojie Xu}{}
\author[4]{Antonios Katsianis}{}
\author[1]{Hong Guo}{}
\author[3]{\\Xiaoju Xu}{}
\author[2,3]{Yizhou Gu}{}
%\author[3]{et al.}{}

%%% Author information for page head. ?¨¹?§Ö????????
%%% ??????????????, ??????????author???
\AuthorMark{Lu Y}%\authorcr????????

%%% Authors for citation. ????????§Ö????????
%%% ??????????????, ??????????author???
\AuthorCitation{Lu Y, Yang X, et al}

%%% Address. ???
%%%   \address[number]{Address, City {\rm Postcode}, Country}
\address[1]{Key Laboratory for Research in Galaxies and Cosmology, Shanghai Astronomical Observatory, Chinese Academy of Sciences, Shanghai 200030, China}
\address[2]{Tsung-Dao Lee Institute, and Shanghai Key Laboratory for Particle Physics and Cosmology, Shanghai Jiao Tong University,  Shanghai 200240, China}
\address[3]{Department of Astronomy, School of Physics and
  Astronomy, Shanghai Jiao Tong University, Shanghai 200240, China}
\address[4]{School of Physics and Astronomy, Sun Yat-sen University, Zhuhai Campus, Zhuhai 519082, China}

%%% Abstract. 
\abstract{We combined data from the Sloan Digital Sky Survey (SDSS) and the Arecibo Legacy Fast ALFA Survey (ALFALFA) to establish the HI mass vs. stellar mass and halo mass scaling relations using an abundance matching method that is free of the Malmquist bias. To enable abundance matching, a cross-match between the SDSS DR7 galaxy group sample and the ALFALFA HI sources provides a catalog of 16,520 HI-galaxy pairs within 14,270 galaxy groups (halos). By applying the observational completeness reductions for both optical and HI observations, we used the remaining 8,180 ALFALFA matched sources to construct the model constraints. Taking into account the dependence of HI mass on both the galaxy and group properties, we establish two sets of scaling relations: one with a combination of stellar mass, $({g-r})$ color and halo mass, and the other with stellar mass, specific star-formation rate ($\rm sSFR$), and halo mass. We demonstrate that our models can reproduce the HI mass component as both a stellar and halo mass. Additional tests showed that the conditional HI mass distributions as a function of the cosmic web type and the satellite fractions were well recovered. }   

%%% Keywords. ?????
\keywords{galaxy groups , neutral hydrogen , halos , statistical}

\PACS{98.62.Ai , 98.62.Gq , 98.65.Cw}%?????

\maketitle

%\tableofcontents%?????

%%%%%%%%%%%%%%%%%%%%%%%%%%%%%%%%%%%%%%%%%%%%%%%%%%%%%%%
%%% The main text. ???????
%???????????????????\cref{fig1}
%\twocolumn\onecolumn
%%%%%%%%%%%%%%%%%%%%%%%%%%%%%%%%%%%%%%%%%%%%%%%%%%%%%%%
\begin{multicols}{2}
\section{Introduction}
\label{sec:intro}

According to the current galaxy formation paradigm, dark matter halos form and grow through gravitational instabilities from small perturbations \cite{Planck2016}. Within the potential wells of these halos, gas cools and condenses, whereas galaxies and stars form \cite{van den Bosch2005,Katsianis2017,Zhang2020}. Thus, it is worth investigating the connections among gases, galaxies, and halos because these connections can provide insights into the underlying physical processes that regulate galaxy formation and evolution \cite{Wechsler2018}.
\Authorfootnote

In the past decade, large HI surveys have provided measurements of tens of thousands of galaxies. Such surveys include the HI Parkes All-Sky Survey (HIPASS) \cite{Meyer2004} which involved $\sim$ 5,000 extra-galactic HI sources out to $z \sim 0.04$ and covers the whole southern sky, and the Arecibo Legacy Fast ALFA Survey \cite{Giovanelli2005} which detected more than 30,000 extra-galactic HI sources from $z \sim 0.06$ in the northern sky. In the near future, the next generation of HI surveys, such as the Five-hundred-meter Aperture Spherical radio Telescope (FAST) \cite{Nan2011}; the on-going Australian SKA Pathfinder (ASKAP) survey; the Wide-field ASKAP L-Band Legacy All-Sky Blind Survey (WALLABY) \cite{Koribalski2012} and the Westerbork Northern Sky HI Survey (WNSHS) \cite{Duffy2012}, will be sufficiently sensitive to detect fainter HI emissions at higher redshifts. Based on these surveys, the global properties of these HI sources, including the HI mass functions of local galaxies, were successfully constrained \cite{Zwaan2005, Martin2010}. Apart from global properties, many extensive studies have been conducted to establish correlations between the HI content and various optical properties of galaxies, including their morphology, luminosity, size, and star formation activity \cite{Boselli2001, Kannappan2004, Zhang2007, Zhang2009, Catinella2010, Cortese2011, Catinella2012, Wang2015}, see more details in a recent review \cite{Saintonge2022}. 

Among the HI-galaxy correlation studies, one particular effort was made to establish scaling relations between the HI mass and various galaxy optical properties. For example, Kannappan et al. \cite{Kannappan2004} found that the HI-to-stellar mass ratio correlates with the optical color of the galaxy, with a scatter of $\sim 0.4$dex. Subsequently, Zhang et al. \cite{Zhang2009} used a linear combination of $i$-band surface brightness and $g-r$ color to estimate the HI fraction of galaxies, and reported the scatter to be $\sim 0.31$dex. Catinella et al. \cite{Catinella2010} pointed out that the $NUV-r$ color is the single best estimator of the HI mass $M_{\rm HI}$, and determined a `gas fraction plane' by employing the stellar mass surface density \cite{Catinella2013}. On the contrary, Toribio et al. \cite{Toribio2011} introduced several optical properties into principal component analysis and found that the best prediction of the expected value of $M_{\rm HI}$ comes from the diameter of the stellar disk ($D_{25,r}$). Meanwhile, Wang et al. \cite{Wang2011} pointed out that the gas fractions are related to the outer disk color. Motivated by this finding, Li et al. \cite{Li2012} added the color gradient to their linear HI estimator and applied
this improvement to the ALFALFA and GASS samples. Denes et al. \cite{Denes2014} provided the simplest HI-galaxy relation, which only involves the magnitude of a galaxy. After applying the above findings to the HIPASS sample, the scatter was found to be $\sim 0.3$ dex. Note however, as pointed out in a recent study by Zu et al.\cite{Zu2020}, HI scaling relations extracted from the direct HI detections from the current shallow surveys suffer from the Malmquist bias. To alleviate this bias, Zu et al.\cite{Zu2020} designed a likelihood model that accounts for the detection probability of ALFALFA and constrained the model parameters using Bayesian inferences. Based on this method, Li et al. \cite{Li2022} further calibrated their HI mass scaling relation and found that it can measure conditional HI mass functions.

In addition to scaling relations based solely on galaxy properties, numerous studies have investigated their environment dependencies \cite{Rasmussen2012, Serra2012, Brown2016, Stark2016}. Through statistical analyses of the HI gas content of member galaxies within clusters like the Virgo and Coma clusters, it has been noticed that most massive groups are deficient in HI, especially toward the center \cite{Solanes2001, Gavazzi2013, Taylor2012, Cortese2008}, while the situation remains unclear in smaller halos \cite{Rasmussen2012}. For a wider halo mass range, environmental effects were examined using statistical samples. Using a control sample constructed from isolated field galaxies with similar stellar masses and redshifts, some studies have concluded that the HI content can be affected by the properties of the host halo/group \cite{Catinella2010, Catinella2012, Hess2013, Yoon2015} or the local density \cite{Fabello2012}. Satellite galaxies in halos of different masses have very different HI gas fractions \cite{Brown2017}. By measuring the total ALFALFA HI mass in the given SDSS galaxy groups, Lu et al. \cite{Lu2020} found that this value is independent of the halo mass in the range $> 10^{11}\msunh$. Such kind of behavior was further confirmed in a subsequent study by Guo et al. \cite{Guo2021} by stacking ALFALFA data cubes of galaxy groups with different halo masses. Consistent findings have been reported in various studies, such as those by Dev et al.\cite{Dev2023}, who explored the relationship between HI mass and halo mass using the GAMA survey and observed a leveling of the HI mass at higher halo masses. Similar results were obtained by Hutchens et al.\cite{Hutchens2023} and Rhee et al.\cite{Rhee2023}.

Taking into account the above-mentioned HI mass dependence on both the galaxy properties and group environment, in this paper, we set out to obtain more generalized scaling relations that contain both of these two components. Here, we used HI sources observed by ALFALFA that were matched to Sloan Digital Sky Survey (SDSS) \cite{York2000} galaxies for our study. The group environment was adopted from the group catalogs constructed from SDSS DR7 \cite{Yang2012} using the halo-based group finder developed by \cite{Yang2005, Yang2007}. To avoid the Malmquist bias that can be induced by the ALFALFA observations, we introduce an abundance matching method that is widely used in the halo occupation distribution framework \cite{Jing1998, Vale2004}, by carefully considering both the SDSS and ALFALFA survey selection effects, to construct the model constraints. Compared to previous studies, our probe has the following advantages: (1) we are able to provide {\it unbiased} scaling relations with a reliable measure of their intrinsic scatters; (2) our model has properly taken into account the environment effect, enabling us to correctly reveal their conditional distribution behaviors both in terms of halo mass and cosmic web type; (3) our model can predict {\it all} the HI sources in halos of different masses, even those beyond the current shallow HI observation survey limits, enabling fairer HI model constraints in galaxy formation theories.

The outline of this paper is organized as follows. In section \S \ref{sec:data} we describe the HI source sample and the galaxy group catalog used in this study. In section \S \ref{sec:correlation} we investigate the relationship between the HI mass and the galaxy properties like stellar mass, color, and star-formation rate, as well as the halo mass of the group in which the galaxies are located. According to these findings, we construct the HI mass scaling relation model and define the model constraints in section \S \ref{sec:model}. In section \S \ref{sec:performance} we present additional tests conducted to evaluate model performance. Finally, the results are summarized in section \S \ref{sec:summary}.

\section{DATA}
\label{sec:data}

In this section, we describe the data used to constrain the HI mass vs. the stellar mass and the halo mass scaling relationship of our work.

\subsection{The ALFALFA HI Sources}

The ALFALFA \cite{Giovanelli2005, Haynes2011} survey is a blind extra-galactic HI survey. It covers approximately 7,000 ${\rm deg^2}$ on the north sky and includes two separate regions: the first region is from $\sim 7.5 \, {\rm h}$ to $\sim 16.5 \, {\rm h}$ RA in the Arecibo Spring sky. The second subset extends from $\sim 22 \, {\rm h}$ to $\sim 3 \, {\rm h}$ RA in the Arecibo Fall sky. Blind observations of the 21-cm emission line is performed using a 305-m single-dish radio telescope at the Arecibo Observatory with an angular resolution of $3.5'$. In this study, we used the final data release (hereafter $\alpha.100$) \cite{Haynes2018}, which contains 31,502 HI sources up to redshift $z \sim 0.06$. Within these HI sources, 25,434 HI were categorized as secure extragalactic sources (labeled as ``Code 1" in the ALFALFA catalog), and 6,068 sources categorized as ``priors" (labeled as ``Code 2" in the ALFALFA catalog). The latter have low signal-to-noise ratios (${\rm S/N} \leq 6.5$) and usually are not as reliable as the former.  

In the ALFALFA catalog, each HI detection is characterized by its angular position in the sky, radial velocity, velocity width ${ W_{50}}$, and integrated HI line flux density ${S_{21}}$. The HI mass (${ M_{HI}}$) is calculated via equation:
\begin{equation}\label{eq:mhi}
{\frac{M_{HI}}{M_\odot}
=
2.356 \times 10^5 D^2_{Mpc}S_{21}\, ,}
\end{equation}
where ${D_{Mpc}}$ is the distance between the sources in ${\rm Mpc}$ and ${S_{21}}$ is the integrated flux in ${Jy \, km \, s^{-1}}$. We note that no correction for HI self-absorption was applied.

\begin{figure*}[!t]
\centering
\includegraphics[height=6.0cm,width=16cm,angle=0]{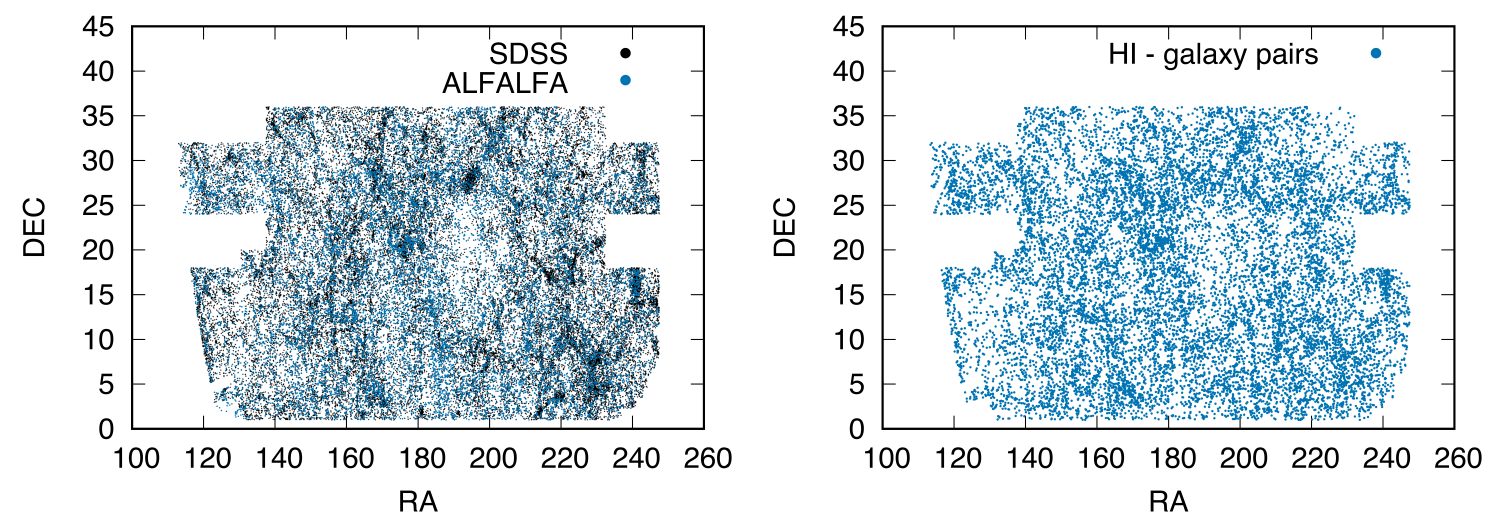}
\caption{Left panel: Parent target distribution. The black dots represent the 69,690 galaxies in the Y07 galaxy group catalog, and the blue dots represent the 20,475 HI sources in the ALFALFA catalog. Right panel: In total, 15,516 HI-galaxy pairs were cross-matched between the Y07 and ALFALFA catalogs within NGC.}
\label{fig:target}
\end{figure*}

\subsection{SDSS galaxy and group catalogs}

In this study, we use the SDSS galaxy group catalogs of \cite{Yang2007} (hereafter Y07), constructed by employing the adaptive halo-based group finder of \cite{Yang2005}, so that, for each galaxy, we have both galaxy properties and relevant group (halo) properties. The parent galaxy catalog is the New York University Value-Added Galaxy catalog (NYUVAGC)\cite{Blanton2005}, here updated to Data Release 7 (DR7) \cite{Abazajian2009, Yang2012}, which contains an independent set of significantly improved reductions. The Main Galaxy group sample was constructed for galaxies in the DR7 complete up to r-band apparent magnitude $r \sim 17.7$. The magnitudes and colors of all the galaxies were based on the standard SDSS Petrosian technique. In this catalog, the $k+e$ corrected luminosities in the SDSS $ugriz$ bands and stellar masses estimated from the SDSS photometry are provided for each galaxy.

For each group included in the Y07 catalogue, the halo mass ${M_h}$ is estimated using two methods. One is based on the ranking of the characteristic group luminosity, while the other is based on the ranking of the characteristic group stellar mass, which is defined as the total luminosity and stellar mass of all group members with $\rmag \leq -19.5$, respectively. Here, the halo mass function used for abundance matching was obtained from \cite{Tinker2008} and adopts a WMAP7 cosmology. The above two halo masses agreed reasonably well with each other, but their differences decreased from $\sim 0.1$ dex at the low-mass end to $\sim 0.05$ dex at the massive end. In this study, we choose the ${M_h}$ based on the ranking of group luminosity. For any groups in which the member galaxies were fainter than $\rmag = -19.5$, the halo masses were estimated according to the stellar-to-halo mass relation for central galaxies obtained in \cite{Yang2012}. In summary, the group catalog used in this study contains 639,359 galaxies in the redshift range $0.01 \leq z \leq 0.20$, distributed among 472,416 groups.

\subsection{HI - galaxy and group counterpart}

In the ALFALFA catalog, among the 31,502 HI sources, 31,158 sources were given optical counterparts by the ALFALFA team, which constitute about $99\%$ of the total number of sources. SDSS images were used to interactively identify the most probable optical counterpart of each HI source. The resolution of the ALFALFA spectral grids was approximately $4'$, while the positions of the HI sources could be determined to an accuracy typically better than $20''$. The identification of optical counterparts is somewhat artificial and is mainly based on information such as color, morphology, redshift, and separation from the HI centroid. After processing each HI source, consistency checks were performed to evaluate any redshift discrepancies or large positional offsets. More details on the search and identification of optical counterparts can be found in \cite{Haynes2011}.

Among the SDSS galaxies identified as optical counterparts, 16,520 galaxies within 14,270 groups were included in the Y07 galaxy group catalog. These HI-galaxy pairs were selected as the target samples for our investigation. Note that to provide better statistic and model constraining power, we include both ALFALFA ``Code 1" and ``Code 2" sources, which signal-to-noise ratios are ${\rm S/N > 6.5}$ and ${\rm S/N \leq 6.5}$, respectively.

In order to reduce the impact of the survey selection effects on our estimations of the HI scaling relations, we construct an SDSS and ALFALFA overlapping galaxy and group sample. We only considered galaxies and groups located in the northern galactic cap (NGC) of ALFALFA, whereas the southern Galactic cap (SGC) was too small to be used. Taking into account the ALFALFA survey depth, we adopted the same redshift range for both catalogs ($0.01 \leq z \leq 0.06$). After the above redshift and geometry cuts, 69,690 galaxies were found in the overlapping region, among which 15,516 ($22\%$) of them are matched with HI counterparts. In the left panel of Fig. \ref{fig:target}, the galaxies of the Y07 and HI sources provided by ALFALFA are displayed by black and blue dots, respectively, while the galaxy-group-HI pairs in the same sky range are represented in the right panel by blue dots.

We notice that there are $\sim$30\% of HI detections appeared in the SDSS and ALFALFA overlap regions, but their galaxy counterparts were excluded from the Y07 galaxy and group catalog. The missing galaxy counterparts are primarily caused by two reasons :
\begin{itemize}
\item They do not possess redshift information.
\item Their $r$-band magnitudes exceeded the magnitude limit of $r\sim 17.7$.
\end{itemize}

\section{Correlations between HI mass and galaxy \& group properties}
\label{sec:correlation}

In this section, we investigate the dependence of the HI mass on different galaxy properties as well as the halo mass. The analysis was carried out using galaxies with available HI detections, i.e., 15,516 HI-galaxy pairs.

\subsection{HI mass ratio dependence on different galaxy properties}
\label{sec:insitu}

Numerous studies have demonstrated that cold gas within galaxies is strongly related to other key galaxy properties. Cold gas has been found to be correlated with galaxy stellar mass ${M_{\ast}}$ \cite{Cortese2011, Catinella2012, Huang2012}, optical color \cite{Kannappan2004} and surface brightness \cite{Zhang2009,Li2012}. In this subsection, we discuss how the HI-to-stellar mass ratio depends on some galaxy properties in our sample. The HI-to-stellar mass ratio is defined as :
\begin{equation}
{f_{HI}=\frac{M_{HI}}{M_{\ast}}}
\end{equation}
 where ${M_{\ast}}$ is the stellar mass of the galaxy.

We present the HI mass ratio $f_{HI}$ of galaxies as a function of 6 key galaxy properties for the HI-galaxy pairs cross-matched between $\alpha.100$ and Y07 in Lu et al. \cite{Lu2020}. We present the dependence of $f_{HI}$ on the r-band absolute magnitude $M_r - 5 \log h$, stellar mass ${M_{\ast}}$, concentration $r_{90}/r_{50}$, the ${g - r}$ color, star formation rate (SFR), and specific star formation rate (sSFR) (defined as the ratio between star formation rate and galaxy stellar mass i.e. $\log ({\rm SFR}/{M_{\ast}})$). Overall, the HI fraction strongly depends on the above galaxy properties, especially the absolute magnitude, stellar mass $M_{\ast}$, color ${g-r}$ and sSFR. Within the relations considered, it is not surprising that the HI fraction strongly depends on the star formation rate (SFR) and specific star formation rate (sSFR) because the HI content provides the fuel to form stars. Next, the HI fraction also strongly depends on the galaxy concentration ${r_{90}/r_{50}}$, where ${r_{90}}$ and ${r_{50}}$ are the radii containing $90\%$ and $50\%$ of the Petrosian flux in the ${r}$-band. There is an obvious break at ${r_{90}/r_{50} \sim 2.6}$, above which a strong anti-correlation between the HI fraction and the galaxie concentration emerges. The above can be explained by the fact that the ${r_{90}/r_{50} = 2.6}$ value is a threshold that divides early-type galaxies from late-type galaxies, which may have already been quenched and have red colors \cite{Strateva2001}.

Among the relationships considered, some are clearly coupled with each other. For example, the luminosity and stellar mass are strongly correlated with each other. Concentration, which is linked to galaxy morphology, can be associated with color. By definition, the sSFR depends on both the SFR and stellar mass. According to the above couplings, we recommend that to account for the overall HI fraction in galaxies and construct a model, only two properties are required. In this study, we recommend two different combinations 1) stellar mass with color and 2) stellar mass with sSFR, which provide $f_{HI}$. The former can be better associated with observations, whereas the latter can be better associated with theory. We note that the above three parameters (stellar mass, sSFR and color) have already been proven to be the primary properties linked to the galaxy HI fraction by other authors \cite{Bell2003,Zhang2009,Li2012}.

\subsection{Total HI mass ratio dependence on halo mass}
\label{sec:survival}

\begin{figure}[H] 
\centering
\includegraphics[height=6cm,width=9cm,angle=0]{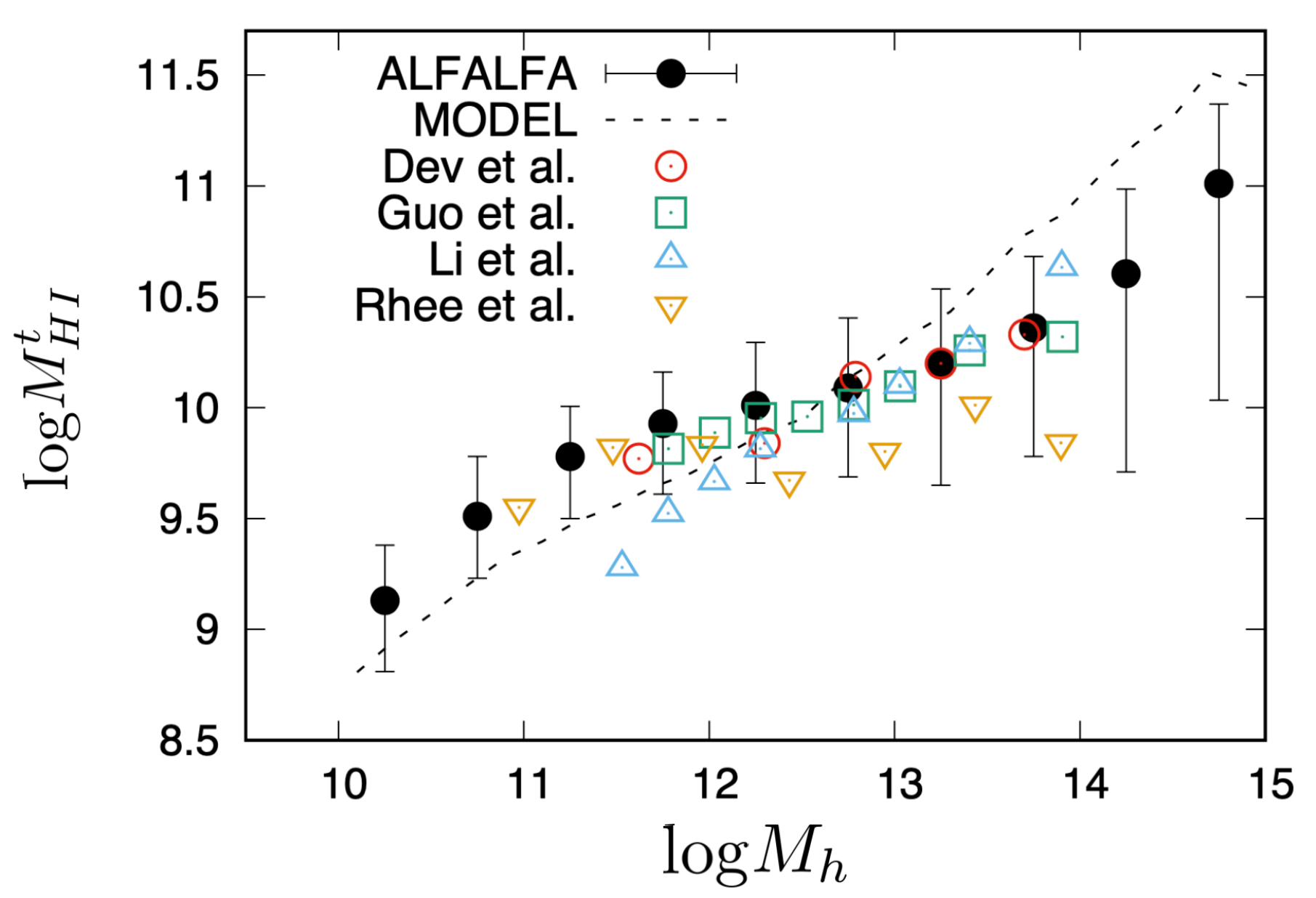}
\caption{Total HI-to-halo mass relationships. Black symbols with error bars represent the median and 68\% confidence level for the ALFALFA sources. The results from other studies are also displayed. The relation given by the proposed HI mass model is also shown for comparison with the black dashed line.}
\label{fig:MhitMh}
\end{figure}

In this subsection, we measure the total HI-to-halo mass relationships within the galaxy groups. Instead of the ${M_{HI}}$ of individual galaxies, we focused on the total cold-gas mass within each halo. This is equal to the sum of the HI masses of all the HI detections of all the member galaxies located within each separate group (halo). The halo mass is a proxy for the halo environment and is included in the Y07 group catalog. In Fig. \ref{fig:MhitMh} we present the total HI mass as a function of halo mass ${M_h}$ for the cross-matched objects in the Y07 and $\alpha.100$ catalogs. The black points with error bars represent the median and $68\%$ confidence level at each halo mass bin, respectively. We see clear dependence of the total HI mass on the halo mass, where the massive halos contain significantly reduced total gas content. Additionally, we incorporated findings from Dev et al.\cite{Dev2023}, Guo et al.\cite{Guo2021}, Li et al.\cite{Li2022}, and Rhee et al.\cite{Rhee2023}, depicted with red, green, blue, and yellow markers, respectively. Our observations align well with those reported by Dev et al. and Guo et al., whereas the trends noted by Li et al. are markedly steeper, and those by Rhee et al.\cite{Rhee2023} appear flatter and more bounded at higher halo masses. In Fig. \ref{fig:MhitMh}, we have also illustrated the correlation according to the HI mass model (labeled MODEL) developed in this study, which is shown as a black dashed line. This model addresses the impact of not including HI-deficient galaxies caused by ALFALFA selection biases.

Due to the survey flux/magnitude limits, the total HI mass may slightly suffer from a certain level of incompleteness, thereby inducing additional scatter in the total HI mass. However, these limits are not directly associated with the halo mass; rather, the related dependence is reserved. In terms of the total HI mass as a function of the halo mass, as shown in \cite{Guo2020, Lu2020}, we see a roughly constant HI mass in halos with masses larger than $10^{11.5}\msunh$. These features demonstrate that the HI mass component is closely related to both the galaxy properties and the halo mass (and probably on the distance to the halo center as well, although not being probed in this study).

\section{Establish HI mass v.s. stellar mass and halo mass scaling relations}
\label{sec:model}

\subsection{HI mass model for each galaxy}

Having demonstrated the dependence of the HI mass on various galaxy properties and the host halo mass, we proceed to establish two sets of scaling relations which can provide the HI mass. The first set is related to the $g-r$ colors of the galaxies, where we assume the HI mass estimator as follows :
\begin{equation}
\label{eq:scaling_color}
%{\log M_{HI}=
%a \log M_{\ast} + b (g-r) - c \log M_h +d\,.}
{\log M_{HI}=
a - b \log M_{\ast} - c (g-r) - d \log M_h \,.}
\end{equation}
The other set is related to the SFR of galaxies, where we assume the HI mass estimator as:
\begin{equation}
\label{eq:scaling_SFR}
%{\log M_{HI}=
%a \log M_{\ast} + b \log {\rm SFR} - c \log M_h +d\,.}
{\log M_{HI}=
a - b \log M_{\ast} + c \log {\rm sSFR} - d \log M_h \,.}
\end{equation}
In addition to these two sets of scaling relations, we have a parameter $\sigma_{HI}$, describing the amount of HI mass scatter in the logarithmic space of each individual galaxy from the scaling relations. Thus, in total, we obtain $a, b, c, d, \sigma_{HI}$ five free parameters that can be constrained using an abundance matching method, as described in the following subsections. As pointed out in recent studies \cite{Wang2018}, central and satellite galaxies do not show significantly distinct quenching fractions once both their stellar mass and halo mass are fixed; here we do not explicitly separate them as well. As demonstrated in Section \ref{sec:performance}, our model can indeed predict the correct HI source satellite fraction.

\begin{figure}[H] 
\centering
\includegraphics[height=7.0cm,width=8cm,angle=0]{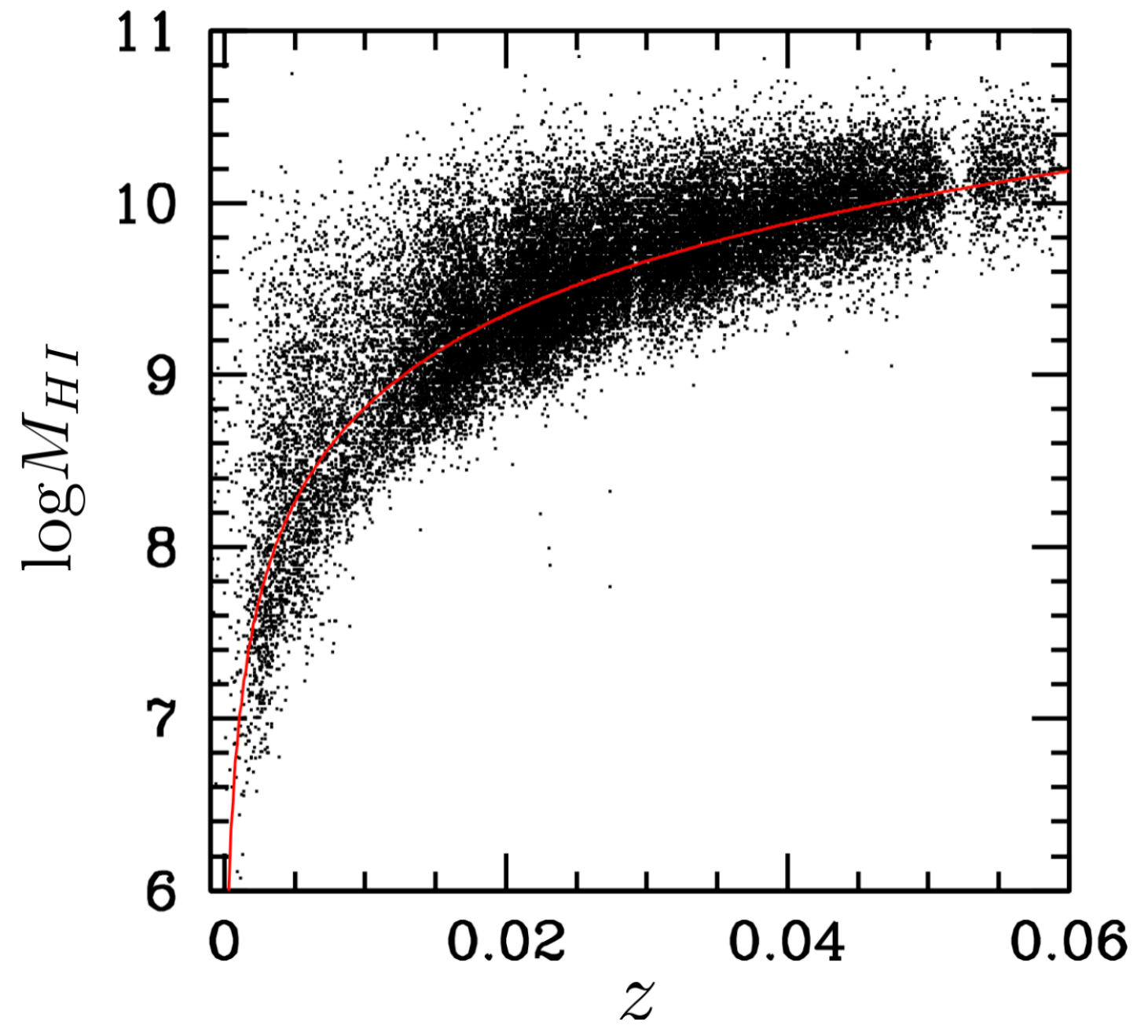}
\caption{Distribution of the ALFALFA HI sources in the mass ${\log M_{HI}}$ v.s. redshift $z$ plane. The red curve roughly corresponds to the HI mass detection limit above which HI detection becomes incomplete.}  
\label{fig:alfaz}
\end{figure}

\begin{figure*}[!t]
\centering
\includegraphics[height=6.0cm,width=16cm,angle=0]{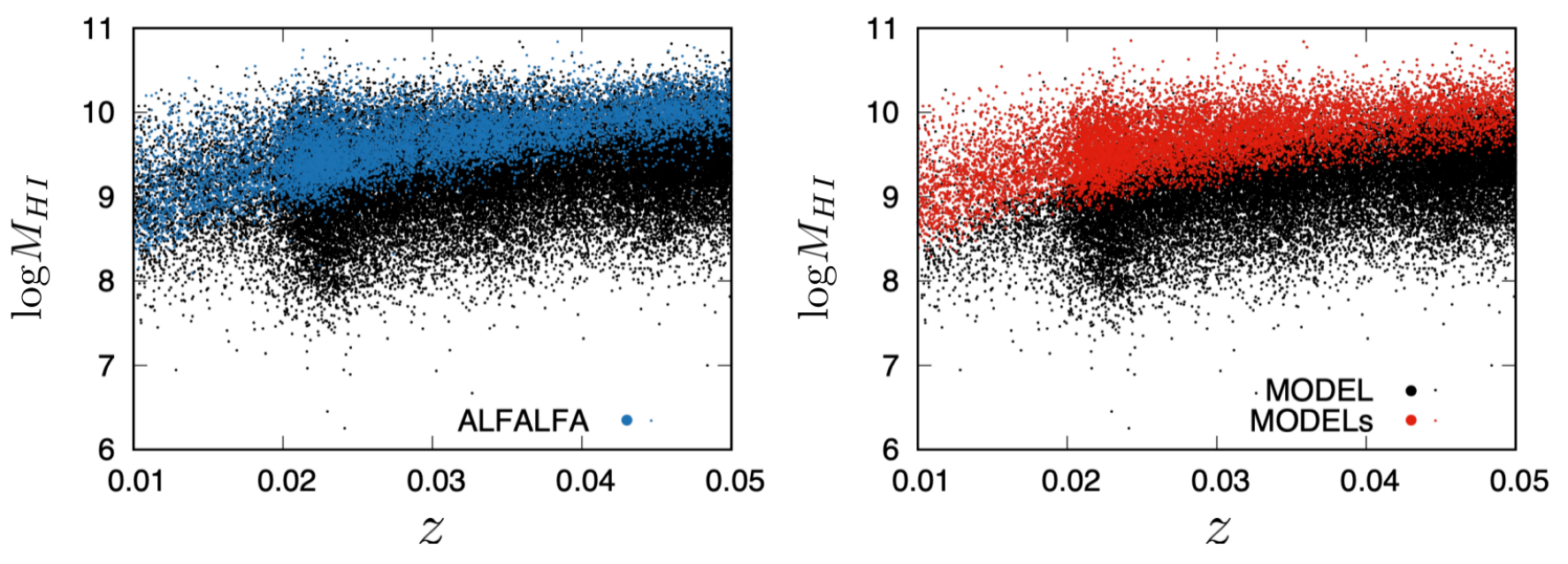}
\caption{Source distribution in an HI mass ${\log M_{HI}}$ v.s. redshift z plane for both ALFALFA observations and model predictions. The black dots in both panels represent the model predictions of all sampled SDSS galaxies. The blue dots represent the distribution of 8,180 sources given by the $\alpha.100$ catalog, while the red dots represent the 8,762 sources predicted by one realization of the best-fitting HI mass scaling relation. Both datasets were subjected to HI observational selection effects. }
\label{fig:number}
\end{figure*}

\begin{figure*}[!t]
\centering
\includegraphics[height=10.0cm,width=17cm,angle=0]{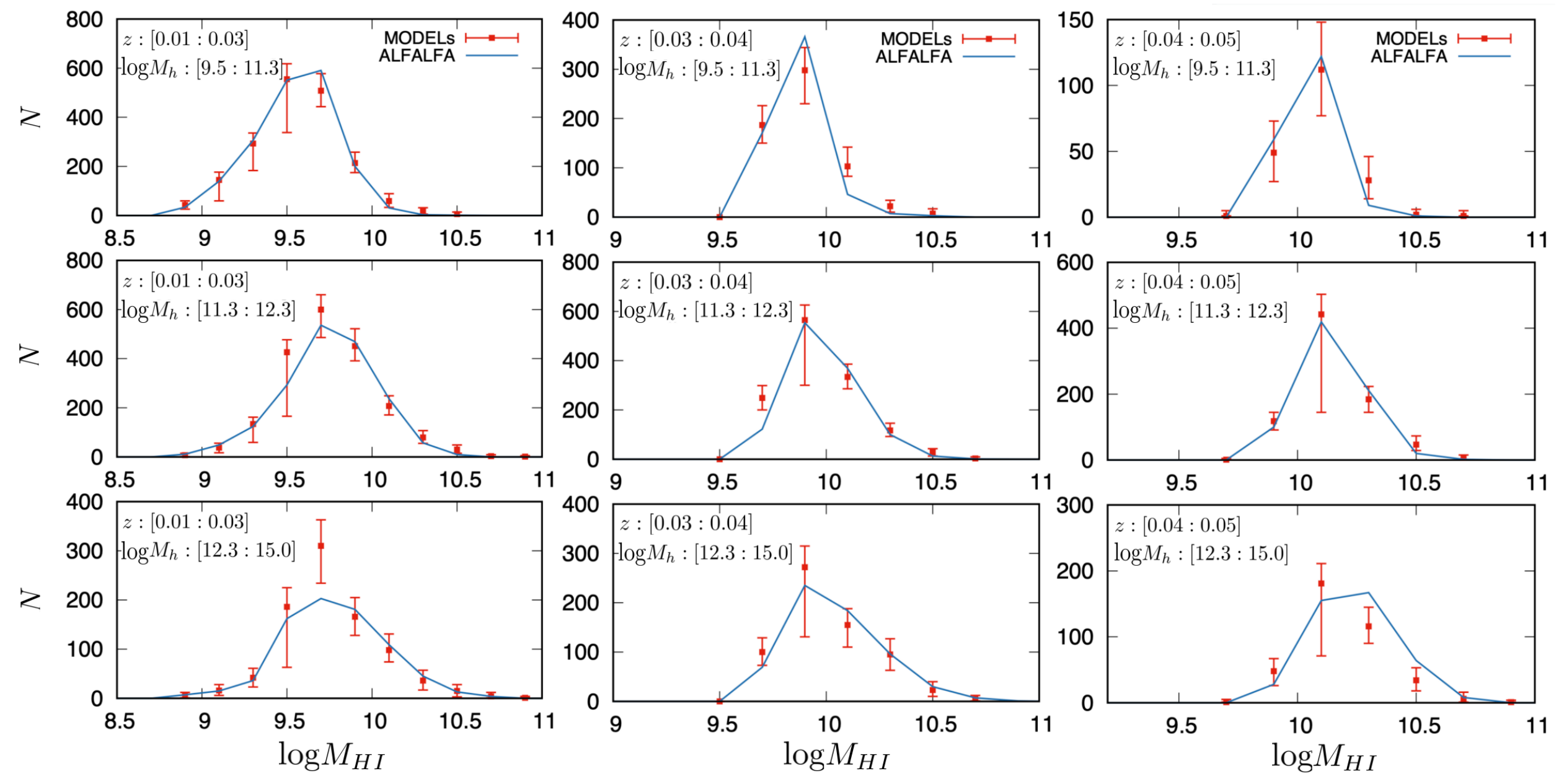}
\caption{Number distribution of HI mass ${\log M_{HI}}$ for ALFALFA observational HI sources (blue solid line) and the model predictions considering the ALFALFA selection effects (red points with errorbars). Each panel corresponds to HI sources within a given redshift and halo mass bin. }
\label{fig:N-Mhi}
\end{figure*}

\subsection{Observational selection effects}
\label{sec:selection}

In this paper, we propose a novel `abundance matching' method to constrain the scaling relations. The abundance matching method has been widely used to establish connections between sub-halos and galaxies \cite{Vale2004}, but it is not widely used to constrain scaling relations. To perform reliable abundance matching model constraints, we need to properly take into account the survey selection effects so that the model data can be matched with observational data.

The ALFALFA is a blind and flux-limited HI survey. The survey depends on both the integrated HI line flux density ${S_{21}}$ and the line profile width ${W_{50}}$, as the detector is more sensitive to narrower line profiles than broader ones at a given ${S_{21}}$. Once a galaxy is assigned an HI mass, the flux density ${S_{21}}$ is calculated using Eq. \ref{eq:mhi}. The line profile width is connected to the intrinsic rotational velocity of the galaxy,
\begin{equation}\label{eq:vrot}
{2 v_{rot} = W_{50} / \sin (\theta)\,.}
\end {equation}
The galaxy inclination angle ${\theta}$ is randomly selected from 0 to 90 degree and assigned to each galaxy in our catalog.
In addition, the rotational velocity correlates with the baryonic mass of a galaxy \cite{McGaugh2012},
\begin{equation}\label{eq:Mbvrot}
{M_b = 47 V^4_{rot}\,.}
\end {equation}
Here, the baryonic mass ${ M_b}$ is the sum of all observed components, including the stellar and gas (HI) masses :
\begin{equation}\label{eq:Mb}
{M_b = M_{\ast} + M_{HI}\,.}
\end {equation}

Using Eqs. \ref{eq:mhi}, \ref{eq:vrot}, \ref{eq:Mbvrot} and \ref{eq:Mb}, we obtain the velocity width of the HI line profile ${W_{50}}$ in ${km \, s^{-1}}$ and integrated HI line flux density of the source ${S_{21}}$ in ${Jy \, km \, s^{-1}}$ for each HI source predicted by our estimator. The ability of a target to be observed or not by a galaxy survey depends on the completeness criterion, which satisfies both the ${ W_{50}}$ and ${S_{21}}$ limits. For ALFALFA, the relationship between the ${S_{21}}$ and ${W_{50}}$ of a source in terms of the signal-to-noise ratio ${\rm S/N}$ of the detection is given by \cite{Giovanelli2005}:
\begin{equation}\label{eq:Slim1}
{S_{21} = \left\{
\begin{array}{ll}
0.15S/N(W_{50}/200)^{1/2}, &  \mbox{$W_{50} < 200$}  \\
0.15S/N(W_{50}/200), &  \mbox{$W_{50} \geq 200$}
\end{array} \right.\,.}
\end {equation}
The above equation gives the expected theoretical survey completeness limit derived from the ALFALFA dataset. According to \cite{Saintonge2007}, the non-Gaussian noise of the automatic signal extractor for ALFALFA is generally above ${\rm S/N = 6.5}$. We assume that for a flux-limited sample from a uniformly distributed population, the number counts follows a power law with an exponent of -3/2. Thus, onset incompleteness can be determined when the data deviate from this form \cite{Haynes2011}. The resulting $90\%$ completeness limit for the ALFALFA sources can be expressed as :
\begin{equation}\label{eq:Slim2}
{S_{\rm 21,lim} = \left\{
\begin{array}{ll}
0.5 \log W_{50} - 1.14, &  \mbox{$\log W_{50} < 2.5$}  \\
\log W_{50} - 2.39, &  \mbox{$\log W_{50} \geq 2.5$}
\end{array} \right.\,.}
\end {equation}

The distribution of the velocity width ${W_{50}}$ versus the integrated flux density ${S_{21}}$ plane based on the $\alpha.40$ catalog \cite{Papastergis2011} shows that the detection limit of the survey is consistent with Eq. \ref{eq:Slim2}. Thus, we use the above method to determine the detectable sources of our mock HI targets generated from the Y07 galaxy catalog. Here, we only include potential HI targets with ${S_{21} > S_{\rm 21,lim}}$. Other blind HI surveys, like HIPASS, estimated catalog completeness as a function of the profile width ${W_{50}}$. The distribution of profile widths ${W_{50}}$ shows a cutoff at ${\rm 30 \kms}$ both for HIPASS and ALFALFA algorithms. We also note that measurements of the velocity width ${W_{50}}$ extend up to ${W_{50} \sim 20 \kms}$, which represents an additional survey limit for the catalog. Here, to comply with the above completeness limit, we also employ a limit of ${W_{\rm 50,lim} = 20 \kms}$ for the simulated HI targets from Y07.

For our subsequent analysis, we only selected mock and ALFALFA observed HI targets with ${S_{21} > S_{\rm 21,lim}}$ and ${W_{50} > W_{\rm 50,lim}}$ in order to comply with the $\alpha.100$ survey limit. Here as an illustration, we present in Fig. \ref{fig:alfaz} the distribution of HI mass (${\log M_{HI}}$) as a function of redshift ${z}$ for each ALFALFA HI source considered in the Y07 and ALFALFA overlap regions. The red line in the plot roughly corresponds to the above HI mass detection limits. Note that in addition to the detection completeness limit, we also observed a significantly reduced number of HI detections at $\sim 0.053$. This feature was found to be caused by the strong RFI generated by FAA radar at the San Juan airport \cite{Haynes2011}.

\begin{figure*}[!t]
\centering
\includegraphics[height=12.0cm,width=12cm,angle=0]{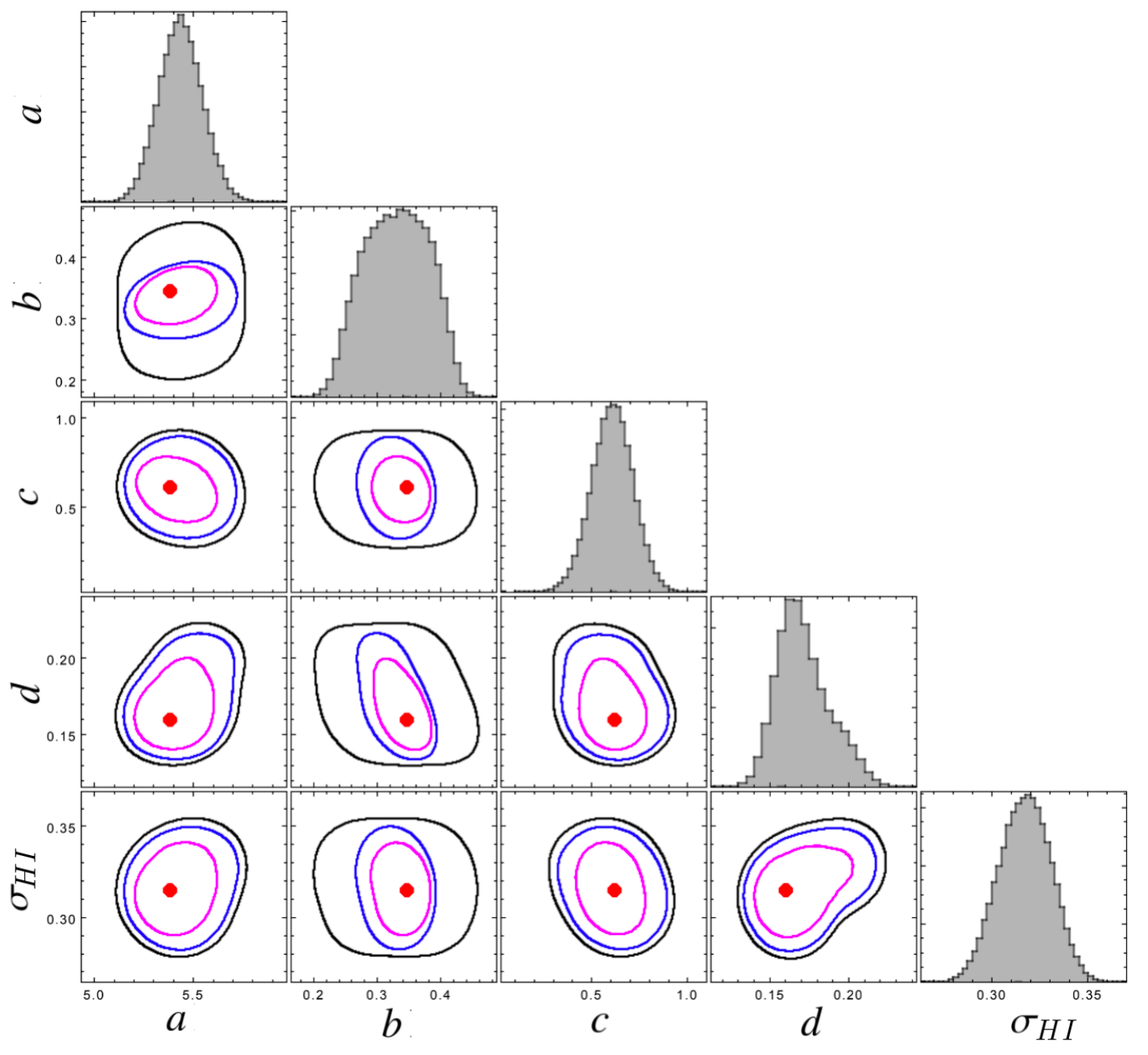}
\caption{The model parameter space distributions given by eq. \ref{eq:scaling_color} with $g - r$ colors. The outer to inner contours correspond to the boundaries that enclose 60\% (black), 10\% (blue) and 1\% (pink) models with the highest likelihood level, while the best-fit parameters are represented by red points.
}
\label{fig:contour}
\end{figure*}

\begin{figure*}[!t]
\centering
\includegraphics[height=10.0cm,width=17cm,angle=0]{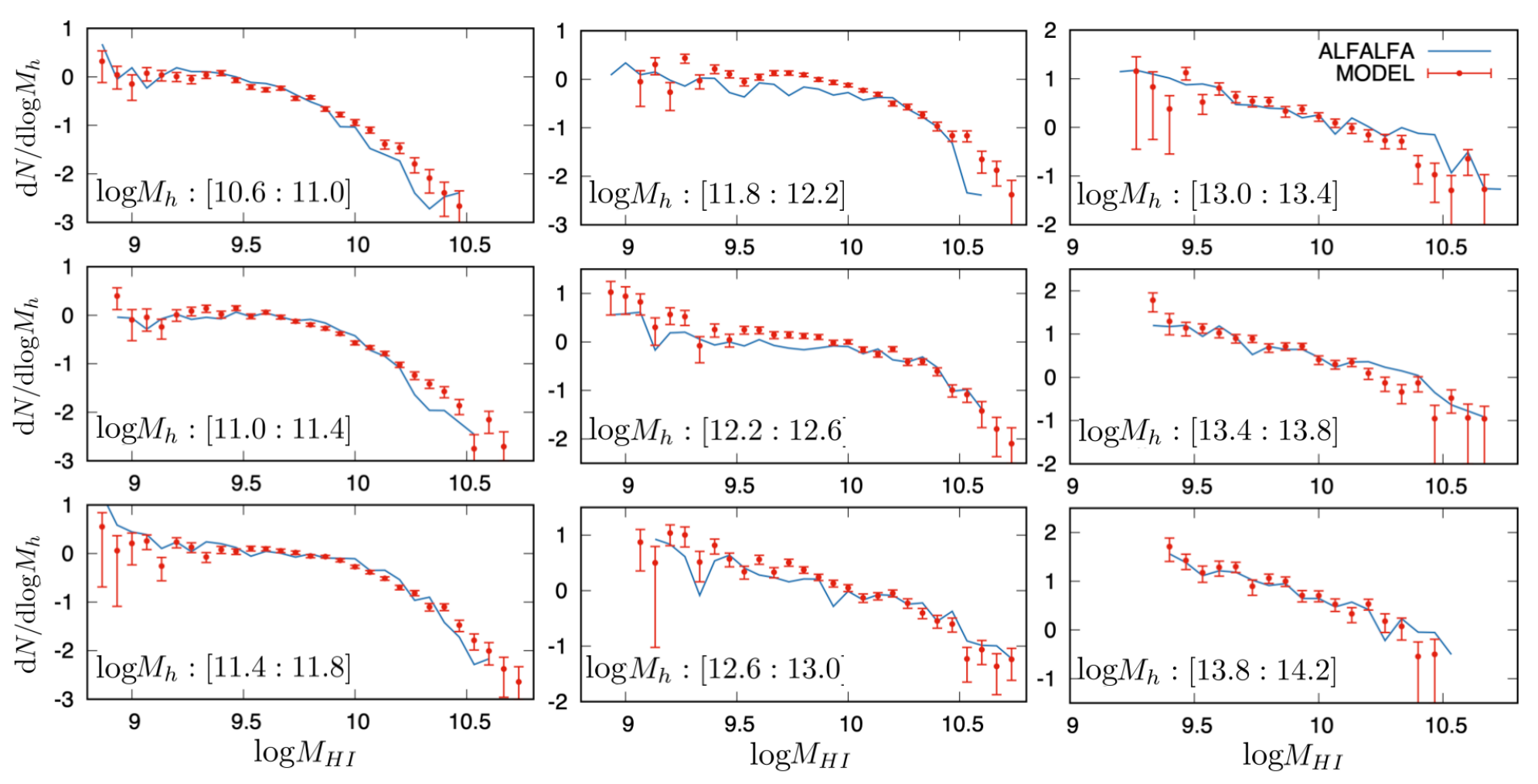}
\caption{Conditional HI mass functions of the observed ALFALFA HI sources (depicted as a blue solid line) alongside model forecasts incorporating the ALFALFA selection biases (shown as red points with error bars). Each panel represents a specific bin of halo mass, as noted. }
\label{fig:HIMF}
\end{figure*}

\subsection{Model constraints}
\label{sec:constraints}

To properly define model constraints, we need to compare the observational data and theoretical predictions under the same selection criteria. To this end, we applied the ${S_{21} > S_{\rm 21,lim}}$, ${W_{50} > W_{\rm 50,lim}}$ and redshift $0.01<z<0.05$ cuts to the observational data, resulting in a total of 8,180 HI-galaxy matched pairs for our subsequent abundance matching model constraints. Note that these sources are matched with the galaxies in Y07 and contain an optical $r$-band magnitude cut $r<17.7$. The distributions of these sources in the HI mass vs. redshift plane are represented by blue dots in the left panel of Fig. \ref{fig:number}.

In order to consider redshift and halo mass dependence, we divided the 8,180 ALFALFA sources into three redshift bins, $0.01<z<0.03$, $0.03<z<0.04$, $0.04<z<0.05$, and three halo mass bins, $9.5<\log M_h<11.3$, $11.3<\log M_h<12.3$, $12.3<\log M_h<15$, respectively. We show in Fig. \ref{fig:N-Mhi} the corresponding HI mass distributions of these sources (blue lines). Each panel corresponds to a redshift bin and a halo mass bin, respectively. 

Our model constraints for the 5 best fit parameters contains the following steps: 

\begin{itemize}
\item Our model contains five free parameters with initial values to be: $a = 6.0$, $b= 0.5$ and $c = 0.5$, $d = 0.3$, and $\sigma_{HI} = 0.3$, respectively. These values were roughly assigned according to the HI-stellar mass ratios discussed in Section \ref{sec:correlation}. 

\item For a given set of model parameters, we assigned HI masses to the 69,690 galaxies in the Y07 and ALFALFA overlapping regions, as shown in Fig. \ref{fig:target}. Here, we use only the HI predictions of 51,292 galaxies located within the redshift range $0.01<z<0.05$ for our model constraints.

\item For each galaxy, we check whether its velocity width $W_{50}$ and integrated flux density $S_{21}$ satisfy the following selection criteria: ${\rm S_{21} > S_{21,lim}}$ and ${\rm W_{50} > W_{50,lim}}$.
 
\item After applying the above selection criteria, the likelihood function value can be calculated by matching the rank of the HI masses of the 8,180 ALFALFA sources and the corresponding survived mock HI sources in the nine redshift and halo mass bins simultaneously. Here, we do not use the number counts as our model constraints; rather, we consider the HI masses of all the ranked sources in each redshift and halo mass bin. We assign each rank galaxy with the same weight and adopt the HI mass uncertainty $\sigma=0.1$ for the model constraints.
\end{itemize}

We used a Monte Carlo Markov chain (MCMC) to explore the likelihood function in the multidimensional parameter space (see \cite{Yan2003} and \cite{van den Bosch2005} for more details), and then we ran the MCMC 300,000 times to obtain the best-fit parameters. Shown in Fig. \ref{fig:contour} are the parameter space distributions. Different contours correspond to different confidence levels, as indicated in the figure caption. In general, the parameters have uncertainties at the levels 10\%-30\%.

Finally, we draw the best-fit scaling relation from the MCMC parameter chain with the minimum $\chi^2$. For color $g-r$, it is given by :   
\begin{equation}
\label{eq:model_color}
\begin{split}
\log f_{HI}=
5.380 - 0.346 \log M_{\ast} - 0.616 (g-r) \\- 0.160 \log M_h \,,
\end{split}
\end{equation}
with a standard log-normal scatter of $\sigma_{HI} = 0.315$. 

We also investigated the replacement of color $g-r$ by the specific star formation rate ($\log SFR/M_{\ast}$). After repeating the same process and constraints, we obtained another set of best-fit scaling relations as follows :
\begin{equation}
\label{eq:model_ssfr}
\begin{split}
 \log f_{HI}=
6.835 - 0.400 \log M_{\ast} + 0.142 \log sSFR \\- 0.145 \log M_h \,,
\end{split}
\end{equation}
with a standard log-normal of $\sigma_{HI} = 0.311$.  

As an illustration, the right panel of Fig. \ref{fig:number} we present the distribution of HI mass as a function of redshift. Here, black and red dots represent the total sample and 8,762 HI sources that fulfill the survey selection effects. Obviously, the ALFALFA survey selection effects allow only the relatively massive HI sources to remain. Compared to the ALFALFA observations shown in the left panel, the distributions with the same selection effects appear very similar. 

Quantitatively, we compared in each panel of Fig. \ref{fig:N-Mhi} the number distribution of the model predictions for the selected sources (red dots with errorbars) with that of the ALFALFA observations. The error bars were obtained from the 1-$\sigma$ scatter using different model parameters for the MCMC chain. As expected, the model data agreed with the observational data very well, as we used the abundance matching method to construct the model constraints. Note that because we used the HI mass distribution in different halo mass ranges for our model constraints, our model automatically predicted the conditional HI mass functions (in halos of given mass) in agreement with the ALFALFA observations. Fig. \ref{fig:HIMF} displays the conditional HI mass functions derived from the ALFALFA data along with the model's predictions for various halo mass bins. The error bars were calculated through 1000 bootstrap re-samplings. Consistent with expectations, the distribution of the HI source from MODEL closely matched that of the ALFALFA data.

\begin{figure*}[!t]
\centering
\includegraphics[height=12.0cm,width=16.0cm,angle=0]{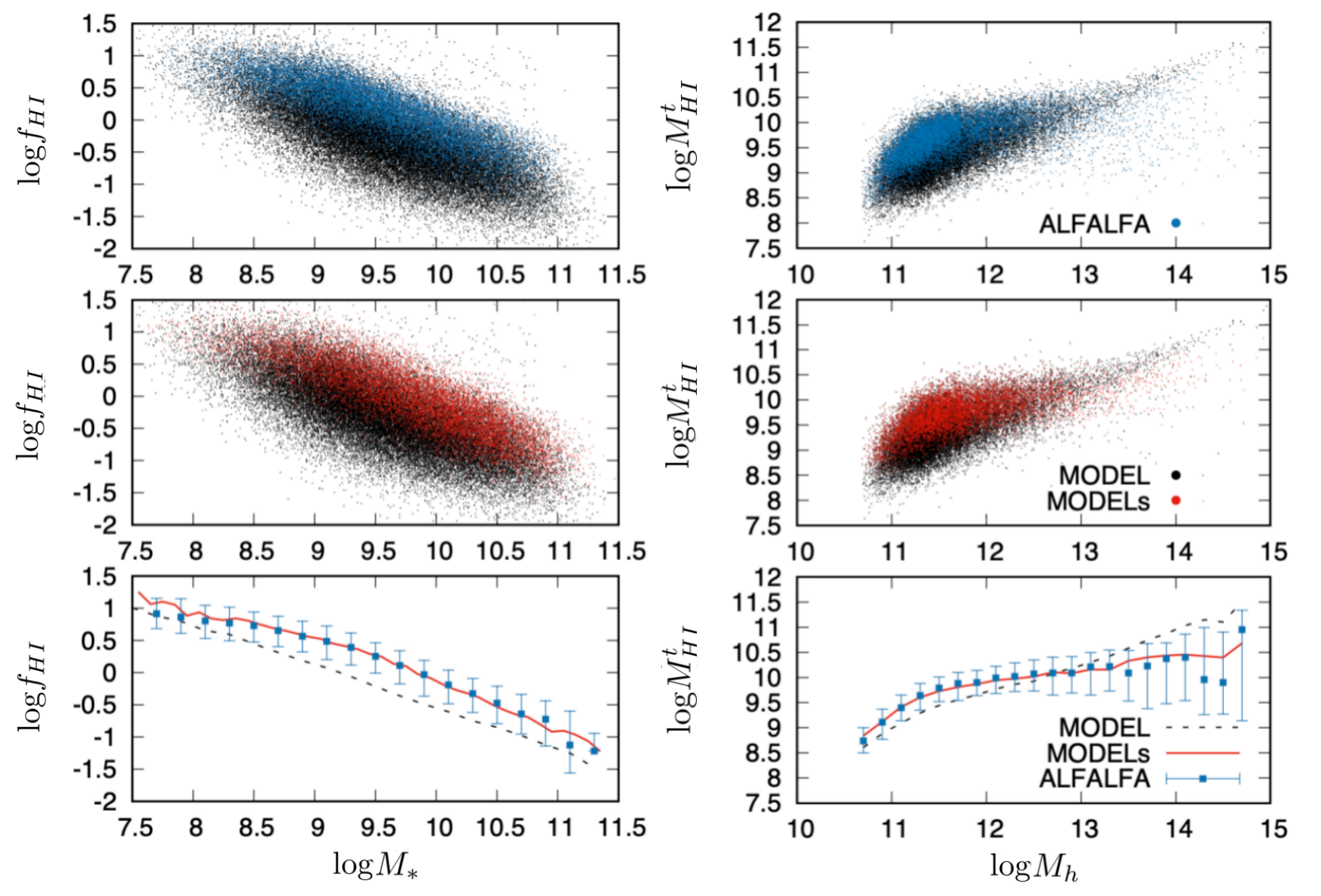}
\caption{The left panels show the relationship between the HI mass fraction and the stellar mass of the galaxy. The top left panel: the observed $M_{HI}$-$M_{\ast}$ relation given by ALFALFA is represented by blue dots, whereas our model predictions of the total sources are represented by black dots. Middle-left panel: Red dots denote the $M_{HI}$-$M_{\ast}$ of our model predictions after applying the same selection effects as in ALFALFA. Lower-left panel: similar to the upper panels, but here we use solid squares with error bars representing the median and 68\% confidence level for the ALFALFA sources. The proposed model is represented by a black dashed line for the total sample and a red solid line for the sample after applying the ALFALFA selection effects. The right panels are similar to the left panels, but they show the total HI mass as a function of the halo mass. }
\label{fig:fracMs}
\end{figure*}

\section{Additional model performance tests}
\label{sec:performance}

In this section, we present further tests on the performance of our two sets of HI mass scaling relations. Since the results for the color-related model and specific star formation-related model are quite similar, here we only present the results for the former.

\subsection[]{HI mass distribution}

We first checked the model predictions of the HI mass fraction ($\log f_{HI}$) as a function of the galaxy stellar mass $\log M_{\ast}$ in the left panels of Fig. \ref{fig:fracMs}. In all three panels from the top to the bottom, the blue color (dots and data points with errorbars) represent the results given by the observations (ALFALFA), while black and red colors (dots and lines) correspond to our model predictions of the total sources (MODEL) and those after applying the ALFALFA observational selection effects (MODELs). 

As shown in the upper-left panel of Fig. \ref{fig:fracMs}, the total HI sources predicted by the MODEL scheme included much more HI-deficient sources than those provided by the ALFALFA survey. Overall, the total HI sources had a larger scatter in $\log f_{HI}$ as a function of stellar mass. After taking into account the ALFALFA observational selection effects, MODELs exhibit quite similar $\log f_{HI} - \log M_{\ast}$ relations as ALFALFA (see middle-left panel of Fig. \ref{fig:fracMs}). In the lower-left panel of Fig. \ref{fig:fracMs}, we can clearly see the median relation provided by the MODELs fits (red solid line) the ALFALFA survey (blue points with errorbars) quite well. However, the relation given by the MODEL scheme (black dashed line) exhibited an obvious shift compared with the observational shift over the entire stellar mass range. This indicates that purely using the ALFALFA-observed sources to constrain scaling relations may suffer from the Malmquist bias.
 
Next, we measured the total HI gas in the galaxy groups. Instead of the ${M_{HI}}$ of individual galaxies, we focus on the ${M_{HI}^t}$, the total cold gas mass within each group/halo. In the three right panels of Fig. \ref{fig:fracMs}, we present the relationship between total HI mass ${M_{HI}^t}$ and halo mass ${M_h}$. Still, the blue, black, and red points/line) represent the results given by the ALFALFA survey, our model prediction of the total HI sources (MODEL), and those after the ALFALFA survey selections (MODELs), respectively. The blue points with error bars represent the median and $68\%$ confidence levels in each halo mass bin, respectively. It should be noted that in this analysis, HI observational selection biases were considered within the ALFALFA survey to exclude incomplete data, as detailed in sec. \ref{sec:selection}.

The total HI mass ${M_{HI}^t}$ increases with the halo mass because larger halos tend to contain more member galaxies and thus more cold gas. However, as shown in the lower right panel of Fig. \ref{fig:fracMs}, we demonstrate that the ${M_{HI}^t}$ and ${M_h}$ relation given by the ALFALFA survey (blue points with errorbars) is not linear. Interestingly, if we only focus on the HI sources that meet the ALFALFA survey selection criteria over a very wide mass range $({\log M_h \geq 11.0})$, the total HI mass changes very slightly, which is in good agreement with the ALFALFA observation data.

However, if we make use of all the model predicted sources, the total HI mass increases with the halo mass (black dashed line) and has a much steeper slope. For halo masses with lower values ($\log M_{h} < 12.5$), the ALFALFA relation is clearly higher than the MODEL. At this halo mass range, due to the survey limit, ALFALFA may eliminate most of these small halos with faint HI signals, which artificially impacts the derived relation (by overestimating the total HI masses at a fixed halo mass). In massive halos ($\log M_{h} > 13.5$), the lack of faint HI sources caused the total HI masses in the massive groups to be underestimated by the survey. Thus, the above selection effects caused some bias in the current ALFALFA observational data. The relationship between total HI and halo masses as predicted by the model is represented similarly by the black dashed line in Fig. \ref{fig:MhitMh}. This discrepancy between the model predictions and observations, not only in ALFALFA but also in other studies, persists. Such discrepancies may be alleviated in deeper HI surveys \cite{Zhang2024}.

To investigate the influence of selection bias on HI-deficient sources, additional verification was performed using xGASS \cite{Catinella2018}. xGASS is a gas-fraction-limited observation of the HI gas content in galaxies, selected only by stellar mass and redshift. This is an extension of the GASS survey that was observed with the Arecibo telescope for a sample of galaxies with redshift $0.01 < z < 0.05$ and stellar mass in the range $10.0 < \log M_{\ast} < 11.5$. The targets were randomly selected from a parent sample of $\sim$12,000 galaxies in the overlap region of SDSS DR6 \cite{Adelman2008}, GALEX \cite{Martin2005} and the ALFALFA survey. Each galaxy was observed with the Arecibo telescope until a significant HI emission line was detected or the HI-to-stellar mass ratio reached an upper limit of $M_{HI} / M_{\ast} \sim 0.015$. The xGASS survey further extends the mass range to $10^{9}\msunh$ and includes galaxies with $9.0 < \log M_{\ast} < 10.2$ and $0.01 < z < 0.02$. Here, we utilize the combined xGASS sample from Catinella et al. (2018) \cite{Catinella2018}, which comprises 1179 galaxies from GASS and xGASS, supplemented with HI-rich galaxies selected from the ALFALFA $\alpha$.70 sample not included in GASS/xGASS. Among the 1179 galaxies, 804 had HI detections. Following the GASS strategy, the following limit is applied to xGASS:

\begin{itemize}
\item $M_{HI} / M_{\ast} > 0.02$ for galaxies with $\log M_{\ast} > 9.7$.
\item Set a constant gas mass limit $\log M_{HI} = 8$ for galaxies with $\log M_{\ast} < 9.7$.
\end{itemize}

\begin{figure}[H] 
\centering
\includegraphics[height=12cm,width=8cm,angle=0]{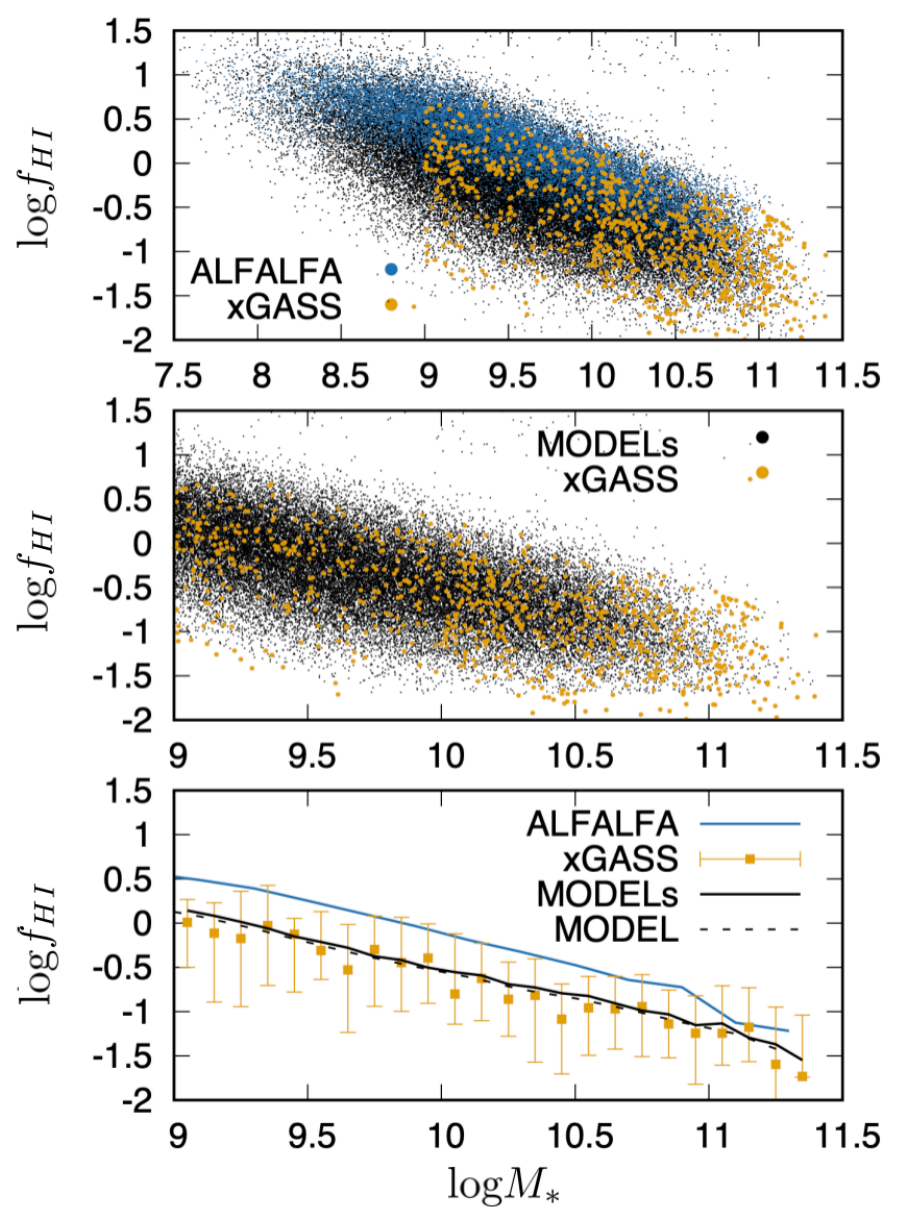}
\caption{As depicted in the left panels of Fig. \ref{fig:fracMs}, this figure illustrates the relationship between the HI mass fraction and the stellar mass of the galaxy. The solid yellow points represent 804 xGASS-detected HI sources. In the middle panel, the black dots labeled as MODELs are selected based on the same standards as xGASS. The data and error bars in the lower panel indicate the median and 68\% confidence level for the xGASS data sources, respectively. In addition, the relationship for the entire sample from MODEL in the bottom panel of Fig. \ref{fig:fracMs} is shown for comparison using the same black dashed line.}
\label{fig:xGASS}
\end{figure}

In the upper panel of Fig. \ref{fig:xGASS}, the 804 xGASS sources are represented by yellow solid dots, and the ALFALFA sources are illustrated in blue for comparison. It is noticeable that the ALFALFA sources predominantly cluster at the gas-rich end, whereas xGASS encompasses more gas-poor galaxies. In the central panel, black dots represent MODELs that adhere to the same criteria as xGASS. This result demonstrates that the proposed model accurately mirrors the general distribution trends of HI galaxies, as evidenced by xGASS. Shown in the lower panel is the median distribution of the HI fraction as a function of stellar mass. Compared with the xGASS results, the ALFALFA results were notably skewed toward a higher HI mass fraction. In addition, we also presented the overall relationship between the HI mass fraction and the galaxy stellar mass as predicted by our model using the black dashed line (the same black dashed line in the bottom left panel of Fig. \ref{fig:fracMs}). Our model prediction agreed with xGASS and MODELs very well, which indicates that xGASS was minimally affected by the Malmquist bias.

\begin{figure*}[!t]
\centering
\includegraphics[height=12.0cm,width=16cm,angle=0]{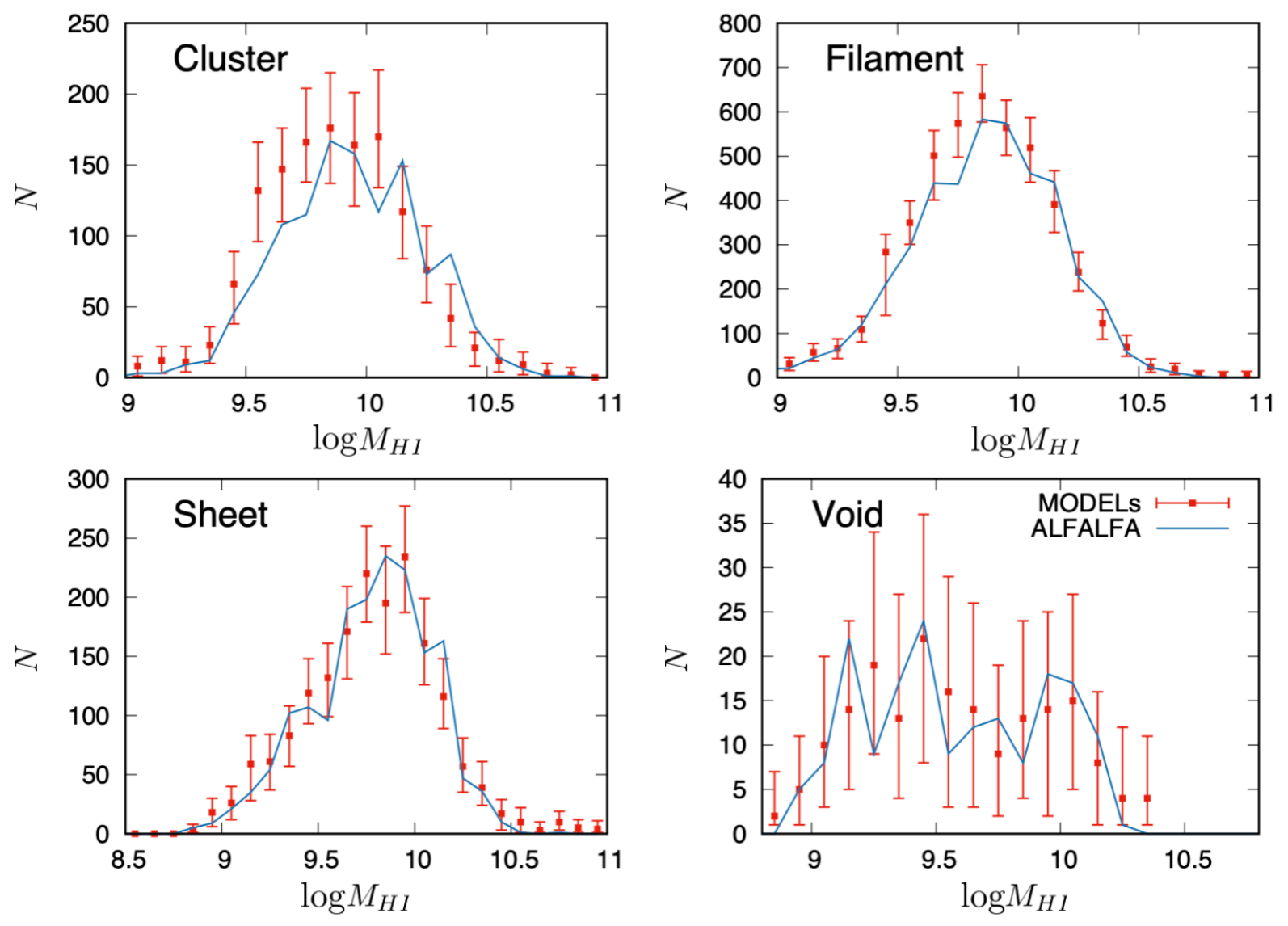}
\caption{Comparison between model predictions and ALFALFA observations of the HI mass distribution in different cosmic web environments, as indicated in each panel. The symbols with errorbars and the solid blue line represent our model prediction under selection effects and the ALFALFA observations, respectively. The error bars are given by 500 bootstrap re-samplings.}
\label{fig:env}
\end{figure*}

\begin{figure*}[!t]
\centering
\includegraphics[height=6.0cm,width=16cm,angle=0]{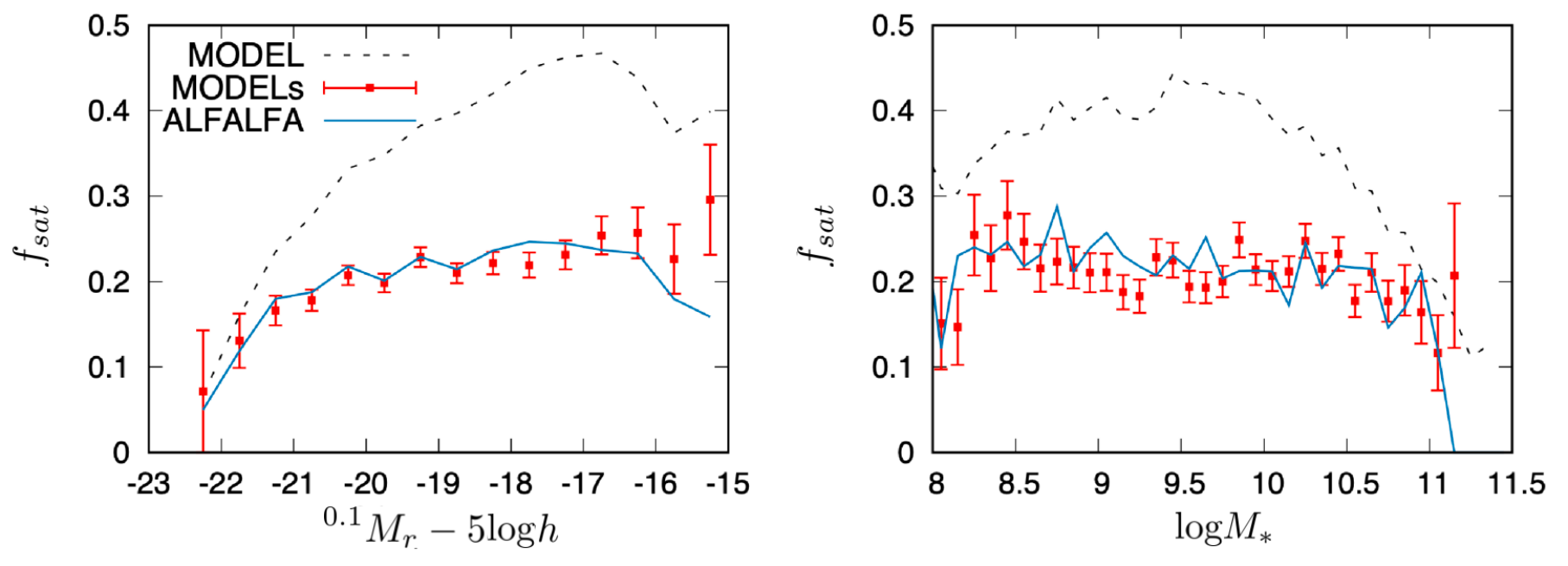}
\caption{The fraction of satellite galaxies ${f_{sat}}$ as a function of absolute magnitude (left panel) and stellar mass (right panel). The dots with errorbars, the dashed black, and solid blue lines represent our model prediction with and without survey selection effects and ALFALFA, respectively. }
\label{fig:fsat}
\end{figure*}

\subsection{Environmental dependence}
\label{sec:environ}

In the previous subsection, we demonstrated the reliability of our HI scaling relations in reproducing the observed dependence on stellar and halo mass. In this subsection, we investigate the HI distribution in different environments to determine whether the HI mass predicted by our model is consistent with the observed mass.

As by construction, our model can provide the correct prediction of the HI mass functions in halos of different masses (see section \ref{sec:constraints}), here we focus on another type of environment indicator, i.e., the cosmic web type. Here, we consider four cosmic web types quantified using a large-scale tidal field and reconstructed from the data using galaxy groups above a certain mass threshold \cite{Zhang2013, Wang2012}. The environment of HI sources can be classified into four classes: cluster, filament, sheet, and void. Fig. \ref{fig:env} describes the distribution of our model predictions, including the selection effects (MODELs) and the ALFALFA HI sources within the different environment classes via the red points with error bars and blue solid line, respectively. As can be seen, in all environments, the distribution of HI masses given by the proposed model can appropriately reproduce the ALFALFA HI source distributions.

\subsection{Satellite fraction of HI sources}
\label{sec:gal}

As a final test, in this subsection, we compare the satellite fractions of galaxies predicted by our model to those obtained from the observed HI-galaxy pairs (Fig. \ref{fig:fsat}). In Fig. \ref{fig:fsat}, the red points with errorbars, the blue solid line, and the black dashed line represent our model predictions taking into account the ALFALFA survey selection effects, the ALFALFA catalog, and the model predictions without selection effects, respectively. The left panel presents the relationship between the satellite fraction and the galaxy absolute magnitude, while the right panel presents the relationship with the stellar mass. We can see that after applying the same selection criteria as for ALFALFA, the satellite fractions presented in this plot are quite consistent between MODELs and the ALFALFA survey. The total satellite fractions were much higher, as demonstrated by MODEL. This indicates that HI-rich sources tend to be located in lower-mass halos as centrals.

\section[]{Summary and discussions}
\label{sec:summary}

In this study, based on the ALFALFA HI observations and SDSS DR7 galaxy and group samples in the same sky region and redshift range, we propose a novel abundance matching method to constrain the extended scaling relations between the HI mass and the galaxy and halo properties. Compared to previous studies, the proposed method and model have the following advantages.

\begin{itemize}

\item Compared to traditional scaling relations purely based on galaxy properties, our extended model properly accounts for the most important halo environment effect, i.e., the halo masses. Here, the halo masses were obtained from the group catalog constructed using a halo-based group finder.

\item Compared to previous direct model constraints using observationally matched HI-galaxy pairs, the abundance matching method does not suffer from the Malmquist bias.

\item In our model constraints, we used flux limit cuts in both the optical and ALFALFA HI observations, which resulted in a large and complete HI-galaxy matched sample for the abundance matching probe.

\item In our model constraints, we separated the 8,180 matched ALFALFA sources into nine redshift and halo mass bins so that the best-fit scaling relations could reproduce the correct conditional HI mass functions in the halos of different masses.

\end{itemize}

For easy application to either observation or theory, we propose two scaling relations that involve the stellar mass, halo mass, the ${g-r}$ color or the $\log sSFR$ of galaxies, as well as a lognormal scatter. By applying our scaling relation models to the optically selected 51,292 galaxies and applying the ALFALFA survey selection effects, we used a total of 8,180 ALFALFA sources to create the model constraints. The best-fit scaling relations related to color and sSFR are provided in Eqs. \ref{eq:model_color} and Eq. \ref{eq:model_ssfr}, respectively.

Additional tests showed that the total 51,292 HI sources did have slight systematic differences from those that fulfill the ALFALFA selection effects. This indicates that purely using the observed ALFALFA sources to constrain scaling relations may suffer from the Malmquist bias. After applying the same selection effects, our model HI sources can reproduce the correct cosmic web dependence (e.g. cluster, filament, sheet and void), as well as satellite fractions.

Finally, we note that in our model, we assumed that the HI mass vs. stellar mass and halo mass scaling relations follow a single log-normal distribution. Limited by the current data quality, we are not yet able to distinguish single log-normal to bi log-normal distribution behaviors. This can be further assessed or improved with future deeper HI and optical surveys (like FAST \cite{Zhang2024} and DESI \cite{Hahn2023, Yang2021, Xu2023}) and more robust modeling of the galaxy-halo connection\cite{Salcedo2022}. We will opt for such a probe in future studies.

\section*{Data Availability}

The data underlying this article will be shared on reasonable request to the corresponding author.
 
%%%%%%%%%%%%%%%%%%%%%%%%%%%%%%%%%%%%%%%%%%%%%%%%%%%%%%%
%%% Conflict of interest. ????????????
%%%%%%%%%%%%%%%%%%%%%%%%%%%%%%%%%%%%%%%%%%%%%%%%%%%%%%%
\InterestConflict{The authors declare that they have no conflict of interest.}
 
%%%%%%%%%%%%%%%%%%%%%%%%%%%%%%%%%%%%%%%%%%%%%%%%%%%%%%%
%%% Acknowledgements. ??§Ý
%%%%%%%%%%%%%%%%%%%%%%%%%%%%%%%%%%%%%%%%%%%%%%%%%%%%%%%
\Acknowledgements{This work is supported by the National Key R\&D Program of China (2023YFA1607800, 2023YFA1607804), the National Science Foundation of China (Nos. 11833005, 11890692, 12141302), “the Fundamental Research Funds for the Central Universities”, 111 project No. B20019, and Shanghai Natural Science Foundation, grant No.19ZR1466800. We acknowledge the science research grants from the China Manned Space Project with Nos. CMS-CSST-2021-A02, CMS-CSST-2021-A03.}

%%%%%%%%%%%%%%%%%%%%%%%%%%%%%%%%%%%%%%%%%%%%%%%%%%%%%%%
%%% Supplements. ????????, ????
%%%%%%%%%%%%%%%%%%%%%%%%%%%%%%%%%%%%%%%%%%%%%%%%%%%%%%%
%\Supplements{}

%%%%%%%%%%%%%%%%%%%%%%%%%%%%%%%%%%%%%%%%%%%%%%%%%%%%%%%
%%% Reference section. ?¦Ï?????
%%% citation in the content using "some words~\cite{1,2}".
%%% ~ is needed to make the reference number is on the same line with the word before it.
%%%%%%%%%%%%%%%%%%%%%%%%%%%%%%%%%%%%%%%%%%%%%%%%%%%%%%%

%%%%%%%%%%%%%%%%%%%%%%%%%%%%%%%%%%%%%%%%%%%%%%%%%%%%%%%
%%% Appendix sections. ??????, ????
%%%%%%%%%%%%%%%%%%%%%%%%%%%%%%%%%%%%%%%%%%%%%%%%%%%%%%%

\end{multicols}
\end{document}